# Transport - and inductive critical current densities in superconducting MgB$_2$

Marc Dhallé, Pierre Toulemonde, Concetta Beneduce, Nicolas Musolino, Michel Decroux and René Flükiger

Département de Physique de la Matière Condensée, Université de Genève, 24 quai Ernest-Ansermet, CH-1211 Genève 4, Switzerland

**Abstract**

The critical current density of four MgB$_2$ samples was measured inductively and for one of them also by transport. Pure phased and dense bulk samples yield a critical current density which in value, as well as in its magnetic field and temperature dependence, is essentially the same as the intra-granular current density measured in a dispersed powder. Also the correspondence between the inductive and transport data indicates that the grain boundaries in the bulk samples are totally transparent for the current. The current-voltage relation becomes rapidly shallow in the vicinity of a depinning line, well below the second critical field. Between the depinning line and the second critical field the material is ohmic and shows a marked magneto-resisitivity, indicative of a flux-flow regime.

## 1. Introduction

In the few months since superconductivity in MgB$_2$ was reported [1], many groups have measured the critical current density and its temperature and field dependence. Most of these data were obtained inductively [2]-[5], with the exception of some early reports on wires or tapes [6]-[8]. The consensus which seems to emerge is that 1) unlike in bulk untextured ceramic superconductors, $j_c(H,T)$ in this material is determined by its pinning properties, and not by weak link effects and 2) similar to high T$_c$ materials, these pinning properties are strongly field dependent, becoming rather poor in modest magnetic field and leading to a depinning line well below the second critical field. In this paper we test these conclusions by 1) comparing inductively measured $j_c(H,T)$ values between different samples and 2) comparing inductive and transport measurements on the same sample.

## 2. Experimental

2.1 Sample preparation

Four samples were studied. The first three are based on a commercial powder (Alfa Aesar, 98% pure). For sample A, the commercial product was ground and dispersed in an epoxy matrix (20% in volume). The purpose of this sample was to measure the intra-granular $j_c$ value, i.e. to eliminate contributions from grain-to-grain currents in inductive measurements. The second sample, B, was hot forged in a uni-axial press at 2 kbar around 1000 °C for 2 h. The third one, C, was sintered in a BN crucible using a 12 mm multi-anvil press under 35 kbar at 950 °C for 1 h, i.e. under similar conditions as those used by Takano et al. [4]. The second family of samples was in-house synthesised, directly from a stoichiometric

Corresponding author : marc.dhalle@physics.unige.ch

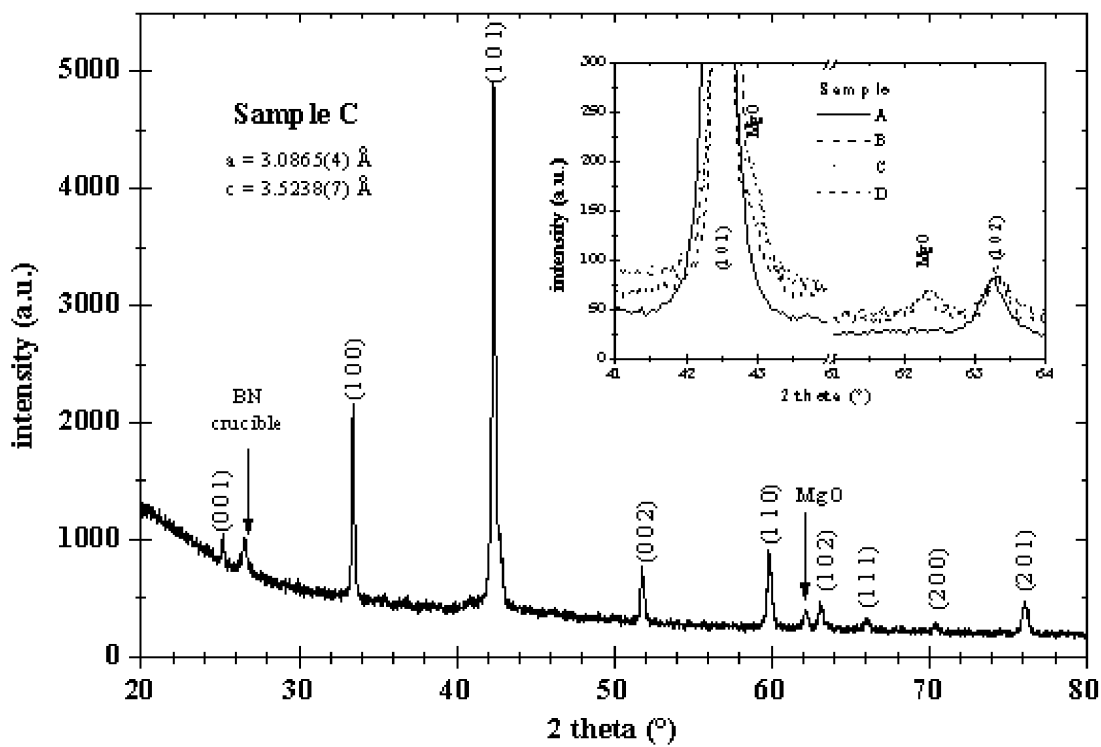

**Figure 1:** X-ray diffraction pattern of the HP-HT sintered commercial $MgB_2$ powder at $\lambda = 1.5418$ Å. The figure inset shows the comparison between commercial $MgB_2$ powder and heated $MgB_2$ pellets: MgO is absent is the former while it appears as small impurity in the second batch of samples.

mixture of metallic Mg and B crystalline powders (supplied by Alfa, respectively 99.8 and 99.7 % pure), at high pressure and high temperature. This preparation route forms an alternative way to the commonly used process based on the reaction between solid boron and a magnesium-rich atmosphere above 650 °C and is particularly well suited for doping purposes. The details will be published elsewhere [9]. The stoichiometric home-made sample used in this study, sample D, was prepared at 850 °C for 1 h, applying a pressure of 45 kbar.

2.2. Structural and microstructural features

The phase content of all the $MgB_2$ samples was checked by powder x-ray diffraction (figure 1) at $\lambda = 1.5418$ Å, $K_\alpha$(Cu). They are nearly phase pure (as far as crystalline phases are concerned, see below for sample D): after heating, under medium (hot-forging) or high pressure, we observe (inset of figure 1) that the dense pellets contain small amounts of MgO (~ 5 %). Its proportion is comparable for the three samples. The MgO content of the untreated commercial powder is under the XRD detection level. The crystallographic lattice parameters of the $MgB_2$ phase are nearly identical in all four samples (Table 1).

| Sample | a (Å) | c (Å) |
|---|---|---|
| A (powder) | 3.079 ± 0.006 | 3.514 ± 0.005 |
| B (hot forged) | 3.083 ± 0.003 | 3.520 ± 0.003 |
| C (HP sintered) | 3.087 ± 0.001 | 3.524 ± 0.001 |
| D (HP synthesised) | 3.083 ± 0.004 | 3.518 ± 0.004 |

**Table 1:** Crystallographic cell parameters of the $MgB_2$ phase in the four samples used in this paper.

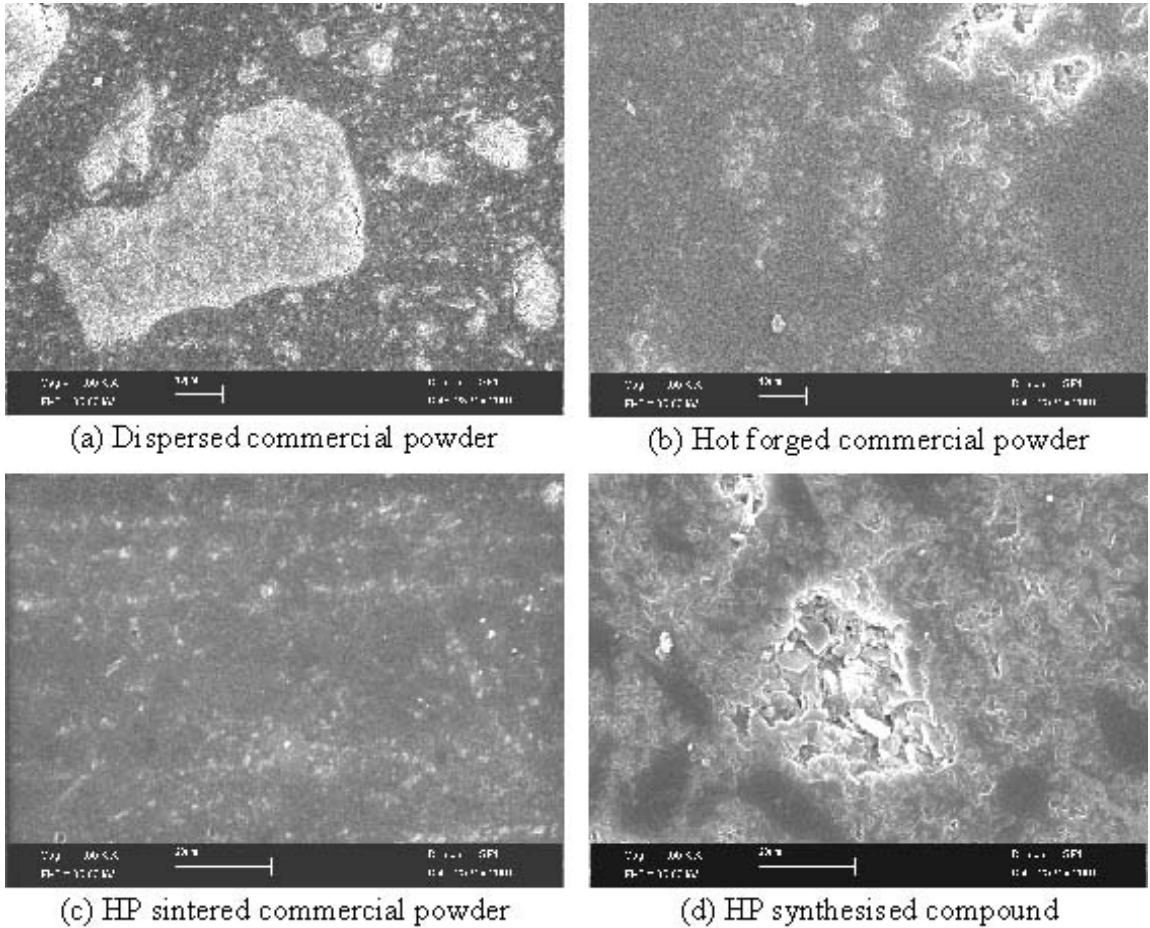

**Figure 2 :** Comparison of SEM images of the four MgB$_2$ samples in terms of density, microstructure and grain size.

The microstructure of the four samples was observed using a Leo-438VP scanning electron microscope equipped with a Kevex system which allows EDX microanalysis. Figure 2 shows typical secondary electron images of polished sections of all four samples. The dispersion of commercial powder in an epoxy, sample A, (figure 2.a.) shows that the grinding did not break up all the well-connected agglomerates of the small (~1 μm in radius) MgB$_2$ crystallites. A quantitative analysis [10] of the agglomerate sizes found in the dispersion showed them to follow an exponential distribution function $f(R)$, averaging at $R_0 \approx 6.5$ μm in radius:

$$f(R) \cong \frac{1}{R_0}\exp\left(-\frac{R}{R_0}\right). \tag{1}$$

Seeing that these remaining agglomerates survived the mechanical grinding, we will assume them to be sufficiently well connected to enable us to inductively measure the intra-granular $j_c$ value.

Figures 2.b and 2.c show how hot forging (sample B, 2 kbar) and high pressure sintering (sample C, 35 kbar) both yield a dense microstructure, with a slightly more pronounced porosity in the case of hot forging. The compound synthesised from elemental Mg and B (sample D, figure 2.d) is clearly less dense than the commercial MgB$_2$ powder sintered at a comparable pressure, possibly due to the formation of isolated gas pockets at some stage of the formation process. Inside one of these pores, the typical MgB$_2$ grain size can be seen to be around 3-5 µm. The small white grains consist of MgO. The numerous dark regions (~25 % in volume) are amorphous and B-rich, presumably corresponding to not yet fully reacted boron particles.

### 2.3 Inductive measurements

The superconducting transition temperature and transition width were measured both inductively and resistively. The inductive $T_c$ measurements were made in an A.C. susceptometer (8 kHz, 0.1 Oe). Inductive $j_c(H,T)$ values were calculated from the $m(H)$ loops measured with a VSM (figure 3). The magnetic field was swept up to 8.8 T and loops were recorded every 2.5 K between T = 7.5 K and T=35 K.

Initial attempts to measure the as prepared pellets (5 mm in diameter) of samples B, C and D lead to flux jumping at dissipation levels above $P/V \sim 0.1$ W/cm$^3$. Therefore we reduced the lateral sample size and thus the induced electric field $E$ and the corresponding dissipation level $P/V = EJ$. From the hot forged sample B we cut a slab of thickness $d = 400$ µm, which was mounted in the VSM with the field parallel to the $3.3 \times 5$ mm$^2$ large face. Pellets C and D were broken into pieces, selected bits were polished into flat platelets ($0.38 \times 1 \times 1.9$ mm$^3$ for C and $0.33 \times 1.1 \times 1.9$ mm$^3$ for D) and again mounted parallel to the applied magnetic field $H$. The Bean formula used to calculate $j_c$ from the measured irreversible magnetic moment $\Delta m(H)$ was that for an semi-infinite slab in parallel field :

$$j_c = 2\frac{\Delta M}{d} = 2\frac{\Delta m}{V\,d}, \qquad (2)$$

with $V$ the sample volume and $d$ the slab thickness. Apart from reducing the electric field and induced power to manageable levels ($E \approx d/2\ \mu_0\ dH/dt$), these platelets offer a well-defined geometry with a low demagnetisation factor, so that equation (2) can be used with some confidence. We estimate the uncertainty on the $j_c$ values thus obtained to be dominated by the uncertainties on the sample dimensions and to be of the order of 10 %.

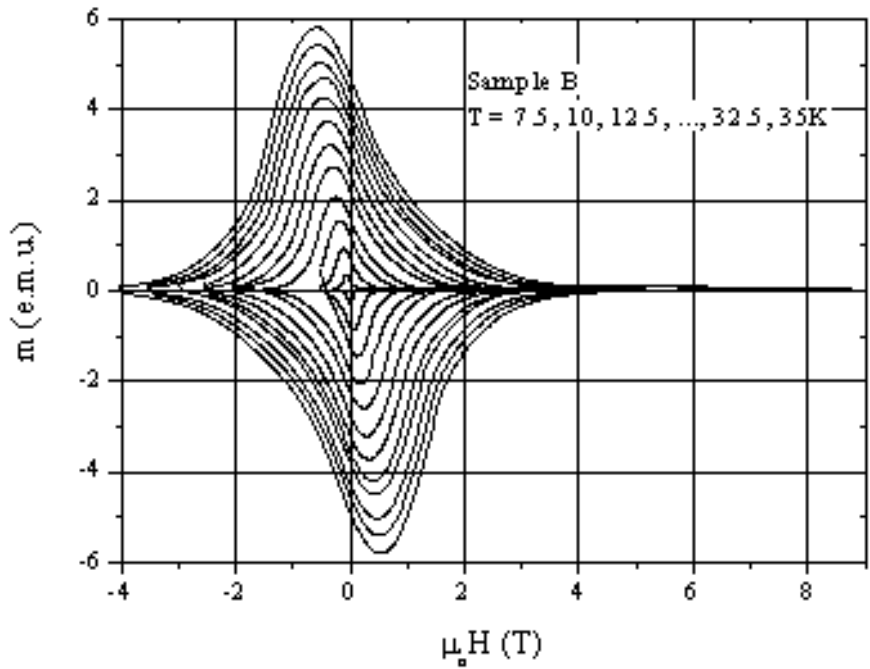

**Figure 3 :** m(H) loops of a thin plate cut from the hot forged MgB$_2$ pellet. Different curves correspond to different temperatures. The magnetic field was applied parallel to the plate.

In order to calculate the critical current density in the dispersed powder, sample A, we assumed the agglomerates to be spherically shaped with radius $R$ and used the formula

$$j_c = 3\frac{\Delta M}{R} = 3\frac{\Delta m}{V\,R}. \tag{3}$$

However, some extra care needs to be taken here. Since the agglomerates have a size distribution $f(R)$ which is not infinitely sharp (see section 2.2), the larger particles will tend to dominate the signal. This can be taken into account by properly weighing the contributions $\Delta m(R) \propto R^4$ coming from agglomerates with different radii $R$ :

$$\Delta m_{total} = N\,\frac{4}{9}\,p\,j_c\,\int_0^{+\infty} R^4 f(R)\,\mathrm{d}R \quad \text{and}$$

$$V_{total} = N\,\frac{4}{3}\,p\,\int_0^{+\infty} R^3 f(R)\,\mathrm{d}R\ ,$$

with $N$ the total number of agglomerates in the sample. Using the exponential size distribution found in the SEM analysis, $f(R) \approx \exp(-R/R_0)/R_0$ with $R_0 \approx 6.5$ µm, we get :

$$j_c = \frac{3}{4}\frac{\Delta m_{total}}{V_{total}\,R_0}, \tag{4}$$

i.e. an effective agglomerate radius $R = 4\,R_0 \approx 26$ µm. The uncertainty on the $j_c$ value will in this case be dominated by the uncertainty on the grain size distribution function $f(R)$, and can be estimated to be around 25 %.

2.4 Transport measurements

The resistive transition of samples B, C and D was measured in a closed cycle cryocooler. For samples C and D we selected small pieces similar to the ones used in the VSM measurements (see above), attached four leads on their circumference using silver paint and measured their resistivity with the Van der Pauw technique. For sample B we used spark erosion to cut small rectangular bars (~ 1 cm × 800µm × 400µm ), which permitted to measure the resistivity and transport $j_c$ in the standard four point configuration.

For the critical current measurements we mounted sample B in a He-flow cryostat. Attempts to tin – or indium solder current leads directly on the polished surface of the bars failed. Pressed indium contacts yielded too high a contact resistance to allow for a reasonable current range for the *I-V* measurements (provoking quenches at $I = 1 - 2$ A, even with the sample submerged in liquid He). Therefore we electrochemically deposited ~ 20 µm of Cu on the sample extremities (figure 4). This allowed for direct tin – soldering of the current leads and delayed quenching of the sample to current amplitudes $I = 30 - 50$ A when cooled in a He gas flow. Interestingly, the dissipation in the sample just prior to quenching could be estimated to be of the order of ~ 0.1 W/cm$^3$, i.e. comparable to the dissipation level which induced flux jumps in the inductive measurements. This suggests that the quenches are not caused by contact heating, but rather by bulk dissipation and the absence of a stabilising matrix in these bare $MgB_2$ samples. During the *I-V* measurements, the temperature was measured on a current lead immediately next to the sample. It remained stable within 0.05 K. The voltage probes were attached 1 mm apart using silver paint. $I_c$ values were determined with a voltage criterion of 1 µV, i.e. at an electric field of $E_c = 10^{-5}$ V/cm (figure 5).

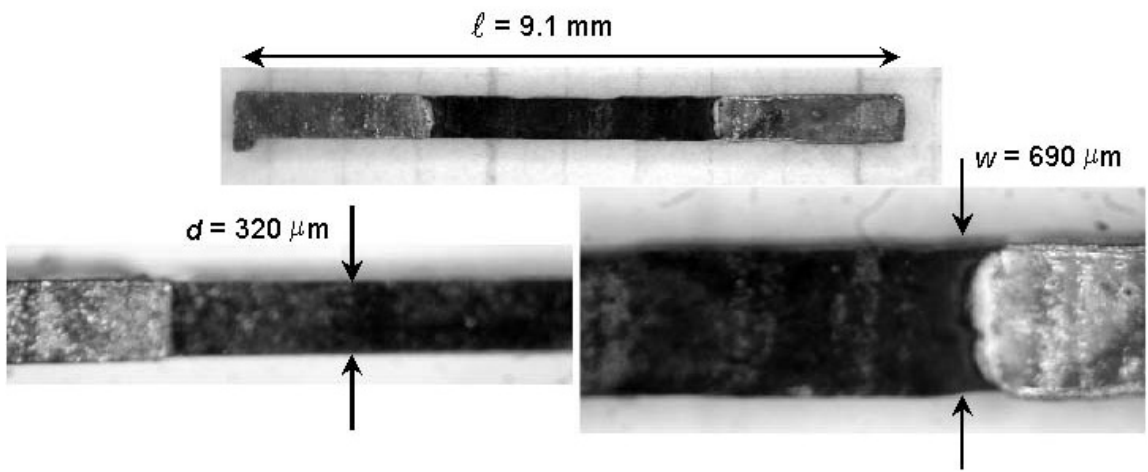

**Figure 4 :** dimensions of sample B used for the I-V measurements

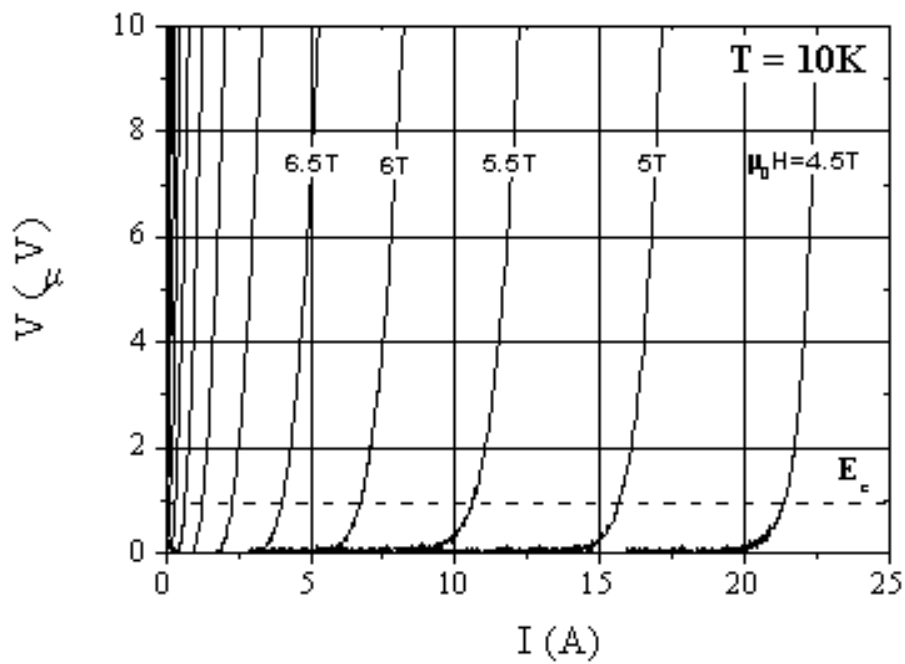

**Figure 5 :** Typical I-V curves obtained with sample B.

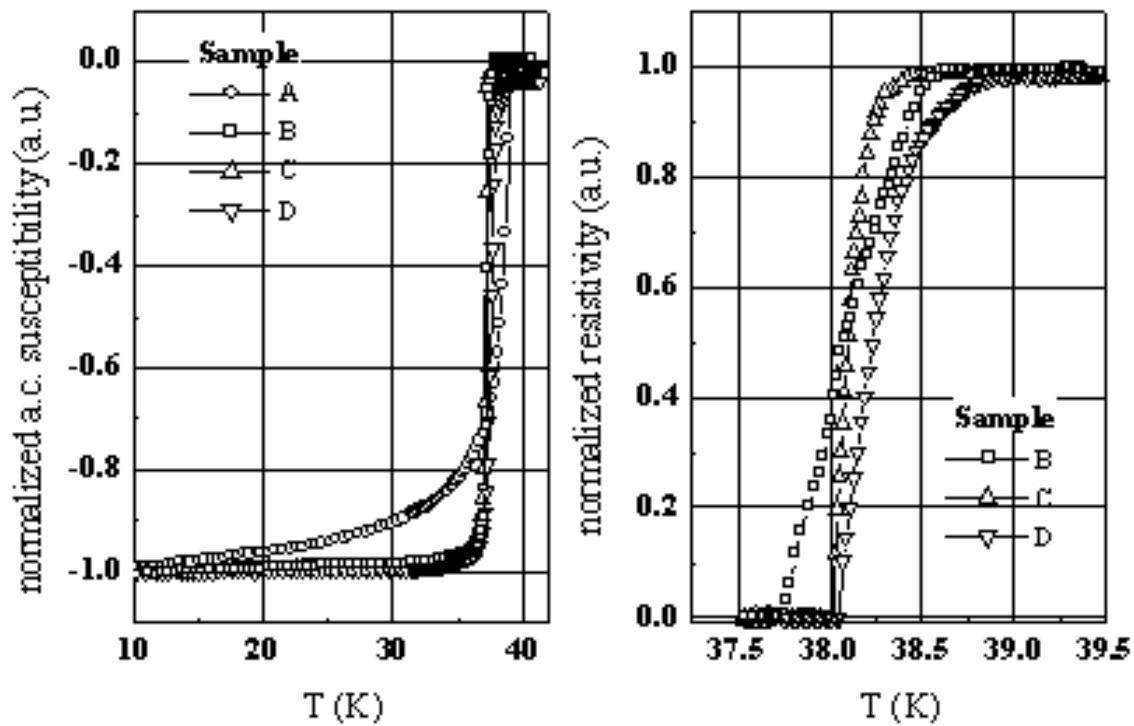

**Figure 6 :** Superconducting inductive (a) and resistive (b) transitions for the different samples studied. The a.c susceptibility was normalised to a low-temperature limit of -1 . The resistivity values were normalised to the normal stare resistivity values reported in table 2.

### 3. Results

3.1 Superconducting transition

The superconducting transition of our samples, measured both inductively and by transport, are shown in figure 6. Figure 6.a shows the normalised A.C. susceptibility of all four samples, while the resistivity measurements for the bulk samples B, C and D are reported in figure 6.b. The residual resistance ratio is approximately 3 for the samples C and D treated in the high pressure anvil, and about 4 for the hot forged sample B. Resistivity values and transition widths (measured between 10 % and 90 % of the transition) are summarised in table 2.

| Sample | $T_{c,onset}$ (K) | $\Delta T_{ind}$ (K) | $\Delta T_{res}$ (K) | $\rho(40K)$ (µΩcm) |
|---|---|---|---|---|
| A (powder) | 39.1 | 8.6 | - | - |
| B (hot forged) | 38.0 | 0.6 | 0.5 | 7.4 |
| C (HP sintered) | 37.9 | 0.5 | 0.2 | 6 |
| D (HP synthesised) | 38.9 | 1.2 | 0.5 | 16 |

**Table 2 :** Inductive $T_c$, transition widths and normal state resistivity of the four samples.

The synthesised sample D presents a wider superconducting transition and a higher resistivity value in comparison with the others prepared from the commercial powder. Moreover, by applying a magnetic field the broadening of the superconducting transition is smaller for the samples B and C. This may be due to a poorer homogeneity of sample D, in accordance with the SEM observations (Figure 2) which show for this sample numerous inclusions of non-reacted boron, suggesting an incomplete reaction.

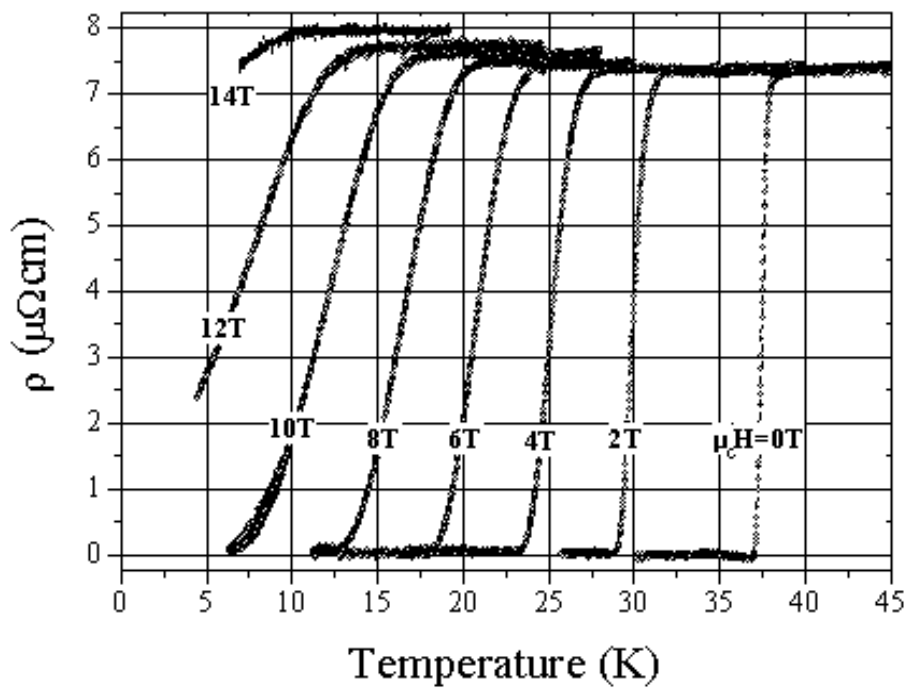

**Figure 7 :** Temperature dependent resistivity of the hot forged sample B in magnetic fields up to 14T.

Fig. 7 shows the resistive superconducting transition of sample B under magnetic fields up to 14 T. The magnetic field was applied perpendicular to the sample surface and the current direction. Note how, just as reported elsewhere for sufficiently low-resistive $MgB_2$ [6], the normal state resistivity exhibits a significant magnetic field dependence for fields exceeding ~ 8 T. Increasing the magnetic field both onset and offset temperatures shift to lower temperature, and the superconducting transition becomes slightly broader. 3.2 Critical current density

3.2 Critical current density

*Comparison between samples*

Figure 8 shows the inductive $j_c$ values of the four samples, measured every 2.5 K between 7.5 and 35 K and plotted against magnetic field. All four graphs in this figure have the same scale. At first sight the values of the critical current densities differ considerably from sample to sample. Furthermore, the intra-granular $j_c$ measured in sample A appears to have a qualitatively different low-field dependence. However, when we estimate the remanent self field trapped in the sample after the field excursion used to measure the *m(H)* loops (indicated in the graphs as dotted lines), we see how these differences arise predominantly due to the different lateral sample sizes. Indeed, the relevant sample size *x* inserted in the Bean model is much smaller in sample A (equation 4; $x = 4R_0 \approx 26$ µm) than in the three bulk samples B, C and D (equation 2; $x = d/2 \approx 200$ µm). As also the trapped field profile is expected to increase in amplitude proportional to the sample size, the field region where the internal magnetic induction due to the trapped flux profile is comparable or larger than the externally applied field (the region above the dotted lines in figure 8) is considerably enhanced in the three bulk samples. The effect of the trapped self field is quite clear-cut in sample B. For fields lower than the self-field, the $j_c(H)$ curves flatten off and $j_c$ becomes nearly independent of the external field. These self-field effects need to be taken into account when comparing the different samples.

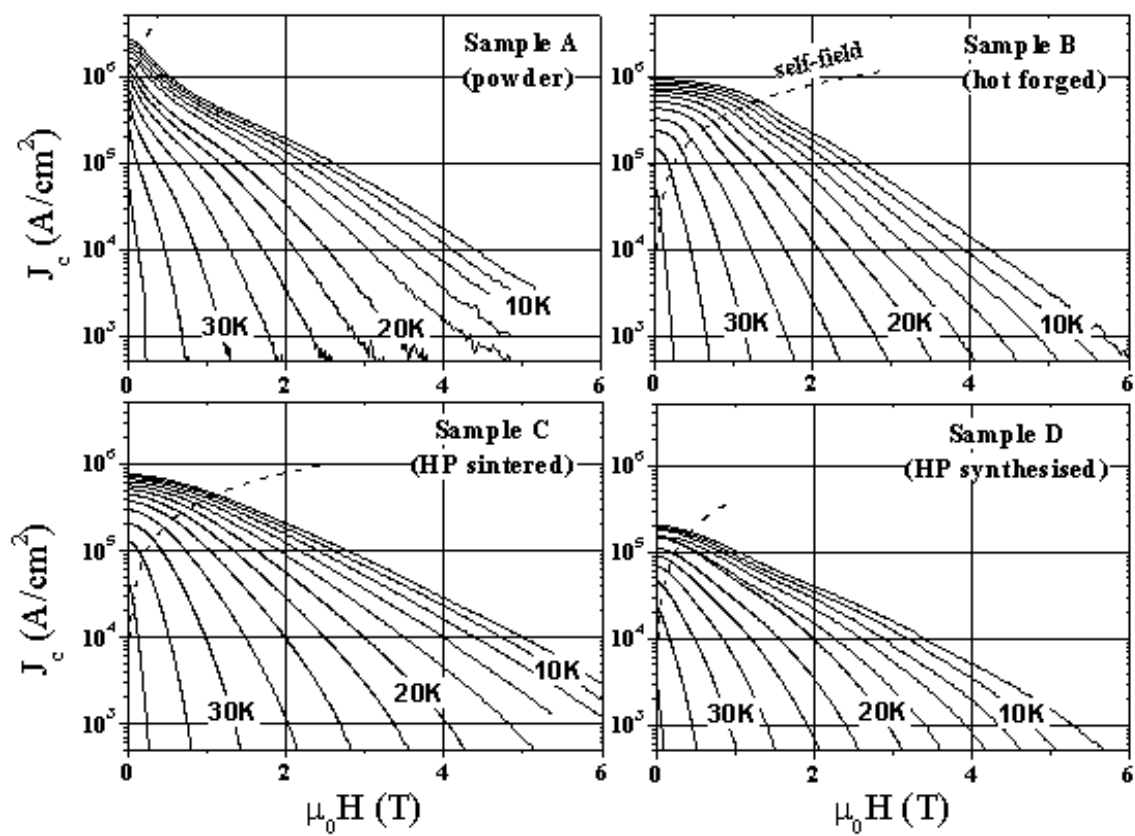

**Figure 8 :** Inductively measured $j_c$ values for the four samples, plotted against applied magnetic field. Different curves correspond to different temperatures between 7.5 and 35K. The dotted lines in each graph correspond to the estimated cross-over between self-field and external-field dominated $j_c(H)$ behaviour.

Figure 9 plots the temperature dependence of $j_c$, inductively measured either in zero external field (left graph) or in an external field of $\mu_0 H = 3$ T (right graph), i.e. well above the estimated self-field. In the self-field case, the $j_c$ value of the hot forged and high pressure sintered samples B and D is about 1/3 of the intra-grain $j_c$ measured on sample A, while sample D yields $j_c$ values which are about 10 times lower than those found in sample A. In a sufficiently high external field, however, these differences become much smaller : sample B carries ~ 90% of the $j_c$ values measured in sample A, while sample C even surpasses these values. Taking into account the absolute errors on these $j_c$ values calculated with the Bean model (see section 2.3), we conclude that there are no significant differences in the inductive $j_c$ values of samples A, B and C. In other words, the bulk critical current density in samples B and C is virtually identical to the intra-granular $j_c$ found in sample A. The temperature dependence of the intra-granular self-field $j_c$ is linear over the whole investigated temperature range :

$$j_c(T) \cong j_c(0K)\left(1 - T/T_c\right), \qquad (5)$$

with $j_c(0K) = 3.3 \pm 0.6$ MA/cm$^2$.

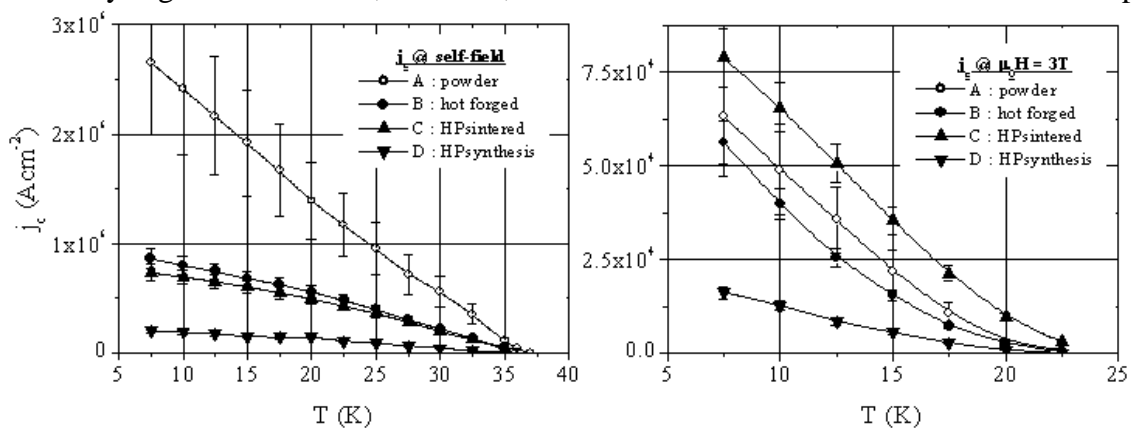

**Figure 9 :** Temperature dependence of the inductive $j_c$ value for the four samples. The graph on the left shows the self-field $j_c$, whereas the graph on the right corresponds to the values measured at $m_0 H = 3$ T, i.e. well above the self-field regime. The error bars reflect the uncertainty on the relevant sample sizes entered in the Bean model.

Sample D carries about 25% of the intra-granular $j_c$ value, which is not too surprising in view of its relatively inhomogeneous microstructure (see section 2.2).

*Comparison between inductive and transport measurements*

Figure 10 compares the inductively measured $j_c$ values of sample B (open symbols) with corresponding transport data (closed symbols). Below $j_c \approx 10^3$ A/cm$^2$, the inductive data start to get noisy, while above $j_c \approx 5\ 10^3$ A/cm$^2$ the transport measurements started to be prone to sample quenches. Nevertheless, the data in figure 10 permit to make some important observations.

First, well above the self field regime $j_c$ decays approximately exponentially,

$$j_c(H,T) \cong j_c(0,T)\exp\left(-\frac{H}{H_p(T)}\right), \qquad (6)$$

with a characteristic decay field $H_p$ which is very similar in the inductive and transport data. The temperature dependence of this decay field $H_p(T)$ is plotted in figure 11, together with the corresponding field measured in the powder dispersion. Both types of measurement yield very similar values for this decay field, so that we can conclude that the field dependence of $j_c$ is essentially the same for transport and magnetisation. Furthermore, it is also virtually identical to the intra-granular $j_c(H)$ dependence as measured on sample A. Just like the self-field $j_c$, the exponential decay field has a linear temperature dependence :

$$H_p(T) \cong H_p(0K)\left(1 - T/T_c\right), \qquad (7)$$

with $\mu_0 H_p(0K) = 0.97 \pm 0.04$ T. Equations (5)–(7) fully describe the intra-granular $j_c(H,T)$ dependence above T = 7.5 K.

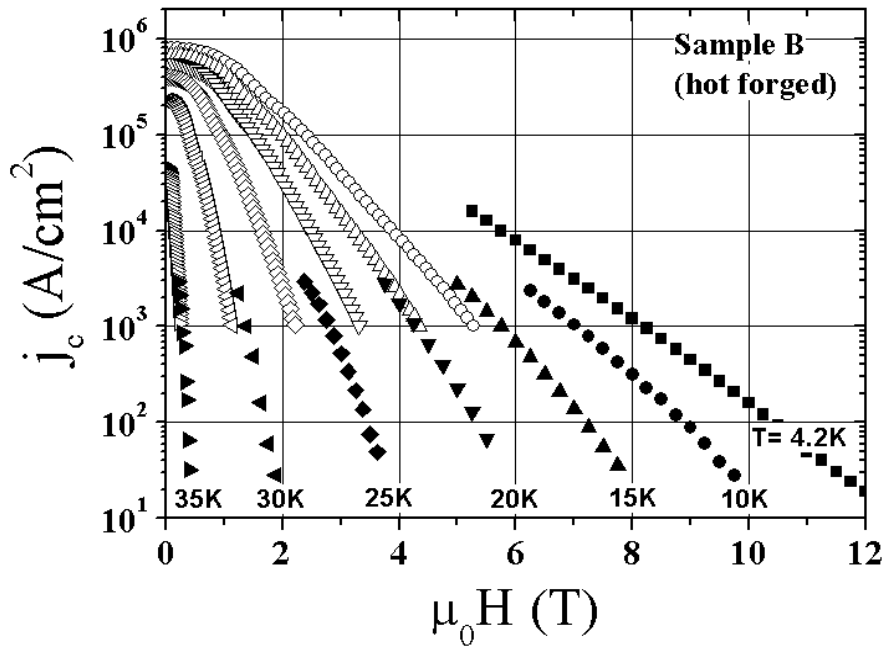

**Figure 10 :** The field dependence of $j_c$ for sample B, measured either inductively (open symbols) or by transport experiments (corresponding closed symbols).

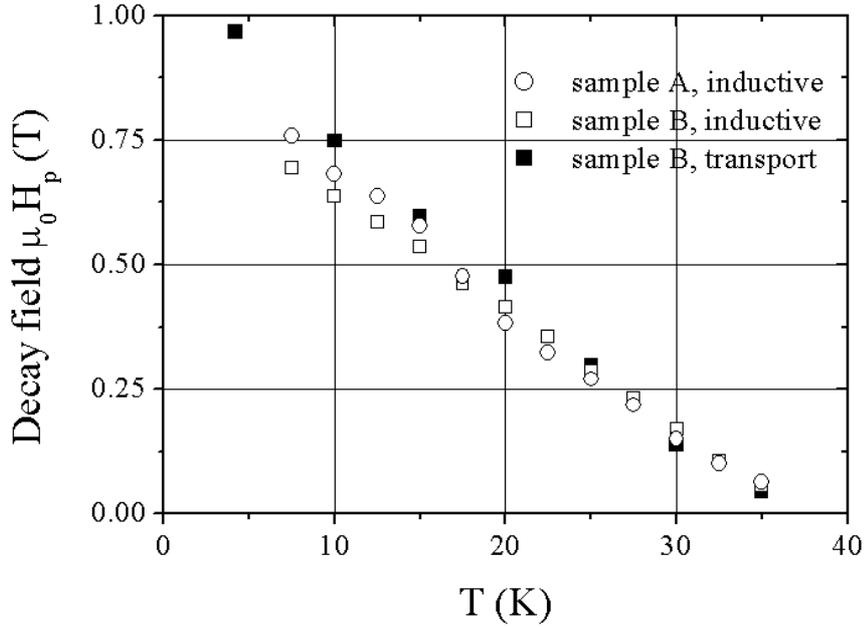

**Figure 11 :** The exponential slope of the $j_c(H)$ dependence well above the self-field regime. The solid symbols correspond to the transport data on sample B, the two data sets represented by open symbols correspond to inductive data obtained on samples A and B.

Second, extrapolating the $j_c$ values measured in the transport experiment to slightly lower fields, we see how they are systematically a factor ~ 2 to 5 higher than the corresponding inductive values. However, this discrepancy can easily be understood as an experimental artefact : at these relatively high fields the *IV* curves become quite shallow (see below), so that "the" critical current density $j_c$ depends significantly on the electric field criterion $E_c$ used in its determination. In the transport measurements this criterion was $E_{c,\text{transport}} = 10^{-5}$ V/cm, whereas in the magnetisation experiments the induced electric field can be estimated to be ~ 1000 times lower : $E_{c,\text{inductive}} \approx 10^{-8}$ V/cm.

Third, both transport and magnetisation data show how $j_c$ becomes negligibly small at magnetic fields well below the second critical field $H_{c2}(T)$, just as measured inductively by numerous other authors [2]-[5] and resistively on MgB$_2$ fibers [6]. Figure 12 plots this experimental depinning field $H^*$ (which we extracted as the field where $j_c(H)$ falls below 500 A/cm$^2$) as a function of temperature. In the same graph we plot the upper critical field $H_{c2}$ (corresponding to the onset of the resistive transitions in field, figure 7). As in other reports [6, 11], the $H_{c2}(T)$ curve shows a positive curvature near $T_c$. At intermediate temperatures the curve is linear and the gradient $dH_{c2}/dT$ is -0.55 T/K. The measured depinning field $H^*(T)$ depends strongly on the experimental conditions : in the transport measurements, $H^*(T)$ is about 70% of $H_{c2}(T)$, while the inductive $j_c(H,T)$ values obtained on sample B yields $H^*(T)$ values which are about half $H_{c2}(T)$. As explained above, this discrepancy is a direct consequence of the relatively low $n$-values of the $E(j)$ relation in this magnetic field range, coupled with the large difference in the electric fields occurring in both type of measurements. Interestingly, the inductively measured $H^*$ values for sample B are nearly identical to the ones found in the powder dispersion of sample A. Since the influence of grain boundaries is minimised in sample A, this confirms the interpretation of the $H^*(T)$ line as a depinning field.

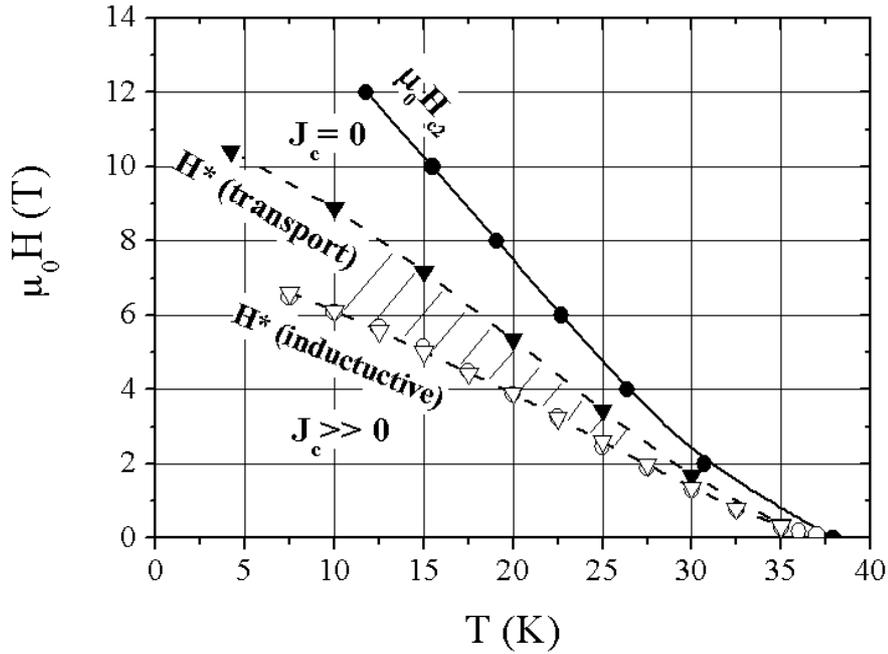

**Figure 12 :** The temperature dependence of the second critical field $H_{c2}$ of sample B, together with the experimental depinning lines $H^*(T)$, measured either inductively (open symbols, samples A and B) or by transport (closed symbols, sample B).

*n-factors*

The logarithmic *EJ* curves measured in the transport experiments tend to show a downward curvature, especially at higher magnetic and electric field levels. Nevertheless, around our electric field criterion $E_c = 10^{-5}$ V/cm they can be reasonably well approximated by a local power-law :

$$\left(\frac{E}{E_c}\right) \approx \left(\frac{j}{j_c}\right)^n \qquad (8)$$

By fitting this relation to our data in the electric field range $5\ 10^{-6} < E < 5\ 10^{-5}$ V/cm, we obtain the n-factors shown in figure 13. Over the whole temperature range between 4.2 K and $T_c$, they have an exponential magnetic field dependence as $H$ approaches $H^*$. Note that a similarly rapid decrease of the pinning potential was also reported to be seen in inductive measurements of the flux creep rate [3] and in A.C. susceptibility measurements [12].

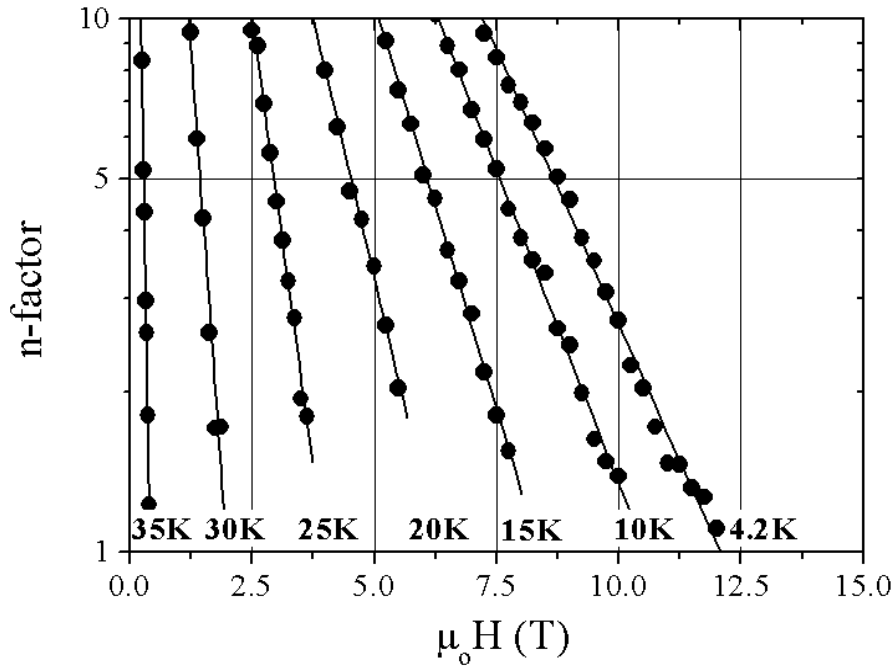

**Figure 13 :** The field dependence of the n-factors describing the logarithmic slope of the electric field – current density relation in sample B.

*Magneto-resistance*

In the magnetic field range above $H^*(T)$ but below $H_{c2}(T)$, the IV curves measured on sample B are linear throughout our experimental electric field window $10^{-6} < E < 10^{-3}$ V/cm. However, their slope changes significantly with magnetic field, i.e. there is a marked magneto-resistance. Figure 14 shows this field dependence of the resistivity plotted at different temperatures. The small arrows above each curve indicate the corresponding $H_{c2}(T)$ value. The strong field dependence of the resistivity between $H^*$ and $H_{c2}$ can be interpreted as a flux flow regime occurring even at very low driving forces $j \times B$, i.e. even in the limit of $j \rightarrow 0$. As such, it confirms how for fields above $H^*(T)$ the flux lines are indeed fully depinned.

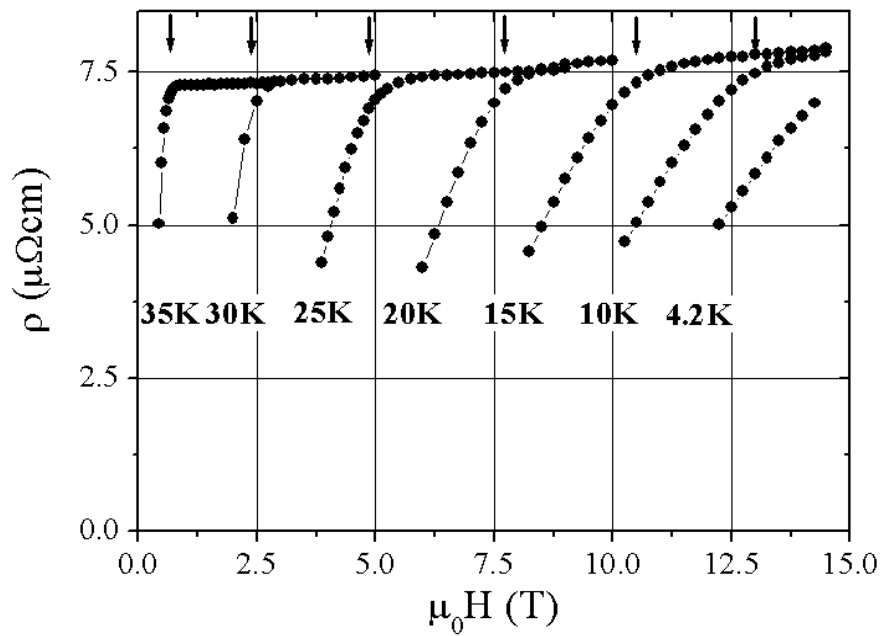

**Figure 14 :** The magneto-resistivity of sample B above the depinning field $H^*(T)$. Different curves correspond to different temperatures. The small arrows indicate the $H_{c2}(T)$ values.

## 4. Conclusions

We measured the critical current density $j_c$ of four $MgB_2$ samples inductively and, for one of them, also by transport.

The inductive measurements indicate that in dense, nearly phase pure bulk samples the macroscopic current density is virtually identical to the intra-granular $j_c$ measured in a dispersed powder. Also the temperature - and field dependence of the bulk $j_c$ is very similar to that of the intra-granular one. The temperature dependence is linear from $T_c$ down to 7.5 K, extrapolating to $j_c(0\,K) = 3.3 \pm 0.6\,MA/cm^2$ for the powder. Apart from self-field effects, $j_c$ decreases exponentially with magnetic field. The exponential decay field $H_p(T)$ has a linear temperature dependence between 7.5 K and $T_c$ and extrapolates to $H_p(0\,K) = 0.97 \pm 0.04\,T$

The transport measurements in high magnetic field on one of the dense and pure bulk samples yields very similar $j_c$ values as the inductive measurements. This confirms that the inductive currents indeed flow coherently throughout the sample, totally unhindered by grain boundaries. Therefore, it has to be flux motion which determines $j_c$ and its temperature and field dependence. The similarity between the $j_c(H,T)$ values measured in the three phase-pure samples indicates that they have comparable pinning properties, suggesting an inherently clean nano-structure in all these samples.

Such an inherently clean nano-structure would also explain the relative weakness of the pinning properties, which gives rise to a depinning line $H^*(T)$ well below the second critical field line $H_{c2}(T)$. The position of this depinning line depends significantly on the experimental method used to determine it. In the vicinity of $H^*(T)$, the $n$-values of the transport $E$-$J$ curves decrease exponentially with magnetic field, indicating a strong field dependence of the pinning potential. Above the $H^*(T)$ line but below $H_{c2}(T)$, these $E$-$J$ curves are ohmic but still field-dependent. Presumably, this corresponds to a flux-flow regime with totally depinned vortices.

**Acknowledgement**

We thank Geraldine Cravotto and Luciano Perez for their help with the electrodeposition of the Cu and Robert Janiec for his assistance with the transport measurements and Aldo Naula for his help with the multi-anvil press..